\documentclass[letterpaper, 10 pt, conference]{ieeeconf}  

\IEEEoverridecommandlockouts                              

\usepackage{graphics} 
\usepackage{epsfig} 
\usepackage{amsmath} 
\usepackage{amssymb}  

\usepackage{verbatim}
\usepackage{graphicx} 
\usepackage{epsfig} 
\usepackage{epstopdf}
\usepackage{amsmath} 
\usepackage{amssymb}  
\usepackage{caption}
\usepackage{color}

\makeatletter
\newcommand{\figcaption}{\def\@captype{figure}\caption}
\newcommand{\tabcaption}{\def\@captype{table}\caption}
\makeatother

\title{\LARGE \bf
Kelly Betting Can Be Too Conservative
}

 \author{\large Chung-Han Hsieh,$^{1}$ B. Ross Barmish,$^{2}$ and John A. Gubner$^{3}$
 	\thanks{\hskip -10pt ${}^1$Chung-Han Hsieh is a graduate student working towards to his Ph.D. degree in the Department of Electrical and Computer Engineering, University of Wisconsin, Madison, WI 53706. E-mail: hsieh23@wisc.edu.}
 	\thanks{\hskip -10pt ${}^2$B. Ross Barmish is a faculty member in  the Department of Electrical and Computer Engineering, University of Wisconsin, Madison, WI 53706. \mbox{E-mail}: barmish@engr.wisc.edu.} 
 	\thanks{\hskip -10pt ${}^3$John A. Gubner is a faculty member in  the Department of Electrical and Computer Engineering, University of Wisconsin, Madison, WI 53706. \mbox{E-mail}: john.gubner@wisc.edu } }

\begin{document}

\maketitle
\thispagestyle{empty}
\pagestyle{empty}

\parindent = 0pt
\begin{abstract}
	Kelly betting is a prescription for optimal resource allocation among a set of gambles which are typically repeated in an independent and identically distributed manner. In this setting, there is a large body of literature which includes arguments that the theory often leads to bets which are ``too aggressive'' with respect to various risk metrics.   To remedy this problem, many papers include prescriptions for scaling down the bet size. Such schemes are referred to as {\it Fractional Kelly Betting}. In this paper, we take the opposite tack. That is, we show that in many cases, the theoretical Kelly-based results may lead to bets which are ``too conservative'' rather than too aggressive. To make this argument, we consider  a random vector~$X$ with its assumed probability distribution and draw~$m$ samples to obtain an empirically-derived counterpart~$\hat{X}$. Subsequently, we derive and compare the resulting Kelly bets for both~$X$ and~$\hat{X}$ with consideration of sample size~$m$ as part of the analysis. This leads to identification of many cases which have the following salient feature: The resulting bet size using the true theoretical distribution for~$X$ is much smaller than that for~$\hat{X}$. If instead the bet is based on empirical data, ``golden'' opportunities are identified which are essentially rejected when the purely theoretical model is used. To formalize these ideas, we provide a result which we call the {\it Restricted Betting Theorem}. An extreme case of the theorem is obtained when~$X$ has unbounded support. In this situation, using~$X$, the Kelly theory can lead to no betting at all.
\end{abstract}

\vskip 3mm

\section{INTRODUCTION}
Kelly betting is a prescription for optimal resource allocation among a set of gambles which are typically repeated in an independent and identically distributed manner. This type of wagering scheme was first introduced in the seminal paper~\cite{Kelly_1956}. Following this work,  many applications and a number of properties of Kelly betting were introduced in the literature over subsequent decades; e.g., see~\cite{Hakansson_1971}-\cite{Algoet_Cover_1988} and~\cite{Maclean_Thorp_Ziemba_2010}. To complete this overview, we also mention more recent work~\cite{Thorp_2006},~\cite{Nekrasov_2014}-\cite{Barmish_Hsieh_2015} and the comprehensive survey~\cite{MacLean_Thorp_Ziemba_2011} covering many of the most important papers.

\vskip 2mm

 In its simplest form, the Kelly  criterion tells the bettor  what is the optimal fraction of capital to wager.  As the optimal Kelly fraction increases, various risk measures can become unacceptably large. In this regard, the optimal Kelly fraction is often characterized as too ``aggressive." To avoid this negative, there is a body of literature dealing with so-called ``fractional strategies.'' Such strategies essentially amount to reduction of the~optimal Kelly fraction so that less capital is at risk on each bet; e.g., see~\cite{Maclean_Ziemba_Blazenko_1992}-\cite{Davis_Lleo_2010} and~\cite{Rising_Wyner_2012}.

\vskip 2mm

In contrast to existing literature, the focal point of this paper is to describe scenarios when the Kelly-based theory may actually lead to bets which are too conservative rather than too aggressive. Our results along these lines are captured in the ``Restricted Betting Theorem,'' its corollaries and generalization given in Sections~5 and~6.  To motivate these results, in the preceding sections, we formally describe the theoretical framework being considered, explain what is meant by ``data-based Kelly betting'' and provide motivating examples which illustrate how overly conservative betting can result. 

\vskip 2mm

With regard to the above, we consider the following scenario: A bettor entertains a sequence of gambles from two different points of view. The first point of view is that of the theoretician who works with a model of the returns as a sequence of independent and identically distributed random variables with a known probability density function. Using the prescription of Kelly for sizing the bet, this bettor arrives at the optimal fraction~$K^*$ of one's wealth which should be wagered on each play. The second point of view is that of the data-based practitioner who makes bets based on an empirically derived probability mass function obtained by drawing samples of the random variable. In this setting, we  describe an example which leads to dramatically different bets for the theoretician versus the practitioner. For this example, we see that a data-based practitioner deems the bet to be  highly favorable and  determines that the optimal betting fraction should be large. However, for this same example, use of the true probability distribution by the theoretician may lead to little or no betting.
 
 \vskip 2mm
 
The main theoretical result in the paper, the Restricted Betting Theorem, is paraphrased for the simplest case, a scalar random variable, as follows: If~$X_{\min}<0$ and~\mbox{$X_{\max}>0$} are respectively the infimum and supremum of points in the support set~${\cal X}$, the optimal Kelly fraction must lie between~$-1/X_{\max}$ and~$-1/X_{\min}$. For the extreme case when the support of the distribution is unbounded both from above and below, this implies the optimal fraction~$K^* = 0$. That is, the optimum is not to bet at all. More generally, when~$X$ is an $n$-dimensional random vector, the
support set~${\cal X}$ imposes a fundamental restriction on size of the optimal bet fraction~$K$ which is described by
$
h_{\cal X}(-K) \leq 1
$
where~$h_{\cal X}$ is the classical support function used in convex analysis. Following the detailed explanation of this result, the final part the paper considers the issue of ``betting frequency'' and how it bears upon the difference of bet sizes for the practitioner versus the theoretician. Finally, in the concluding section, some promising directions for future research are described. 

\vskip 3.5mm

\section{PROBLEM FORMULATION}
In this paper, to make our points about conservatism, we consider one of the  simplest formulations of the problem: The bettor is faced with~$N$ gambles with each individual return governed by an independent and identically distributed~(i.i.d.) random vector~\mbox{$X \in {\bf R}^n$} having probability density function~(PDF) $f_X$. 
On the $k$-th bet, fraction~$K_i$ of one's account value~$V(k)$ is bet on the~$i$-th component $X_i(k)$ of $X$. We allow~$K_i < 0$ so that the theory is flexible enough to allow the bettor to take either side of the bet being offered. For example, if~$X_i > 0$ corresponds to a coin flip coming up as heads, the use of~$K_i = 1/2$ corresponds to a bet of~$50\%$ of one's account on heads and $K_i = -1/2$ corresponds to a bet of $50\%$ on tails. As a second example, in the case of the stock market, allowing~$K_i < 0$ corresponds to short selling;~i.e., when~$X_i(k) < 0$, the bettor wins. In the sequel, we take    
$$
K = {\left[ {\begin{array}{*{20}{c}}
{{K_1}}&{{K_2}}& \cdots &{{K_n}}
\end{array}} \right]^T}.
$$
Then, based on the discussion above, the investment level for the $i$-th bet at stage $k$ is given in feedback form as~\mbox{$I_i(k) = K_i V(k)$} and the associated account value is given by the equation
$$
V(k+1) = V(k) + \sum_{i=1}^{n} I_i(k) X_i(k)
$$
with initial account value $V(0)>0$.

\vskip 2mm

{\bf Admissible Bet Size}:
In the sequel, we let~${\cal X}  \subseteq  {\bf R}^{n}$ denote the support of~$X$ and we require for all~$x \in \cal X$, the admissible~$K$ must satisfy the condition
$$
1+K^T x \geq 0.
$$
The condition above is to assure satisfaction of the survival requirement;~i.e., along any sample path,~$V \ge 0$.
Henceforth, we denote the totality of corresponding constraints above on~$K$ by~${\cal K}$. Now letting~$X(k)$ be the \mbox{$k$-th} outcome of~$X$ for $k=0,1,2,\ldots,N-1$, the dynamics of the account value at stage $k+1$ are described by the recursion
$$
V(k+1) = (1+K^T X(k))V(k).
$$
Then, the Kelly problem  is to select~$K \in \cal{K}$ which maximizes the expected value of the logarithmic {\it growth}
$$
g(K) \doteq \frac{1}{N} \mathbb{E} \left[ {\log \left( {\frac{{V(N)}}{{V(0)}}} \right)} \right].
$$
Using the recursion for~$V(k)$ above and the fact that the~$X(k)$ are i.i.d., we see that the expected log-growth function reduces to
\begin{eqnarray*}
	g(K)
	&=& \frac{1}{N}\mathbb{E}\left[ {\log \left( {\prod\limits_{k = 0}^{N - 1} {\left( {1 + K^T X\left( k \right)} \right)} } \right)} \right]\\ [2pt]
	&=& \frac{1}{N} {\sum\limits_{k = 0}^{N - 1} \mathbb{E} [{\log \left( {1 + K^T X\left( k \right)} \; \right)]} } \\[2pt]
	&=& {\int_{\mathcal{X}} {\log } (1 + K^T x){f_X}(x)dx}
\end{eqnarray*}
which is readily shown to be a concave function of $K$.
Subsequently, when the constraint~$K \in {\cal K}$ is included, we seek to find the optimal logarithmic growth
$$
g^* \doteq \max_{K \in {\cal K}}g(K)
$$
 and we denote a corresponding optimal element by $K^*$.
 
 \vskip 3.5mm
 
\section{BETTING BASED ON DATA VERSUS THEORY}
When Kelly betting is used in practice, it is typically not the case that a perfect probability density function model~$f_X(x)$ for the random variable~$X$ is available.  The practitioner obtains a number of data samples~$x_1,x_2,\ldots,x_m$ for~$X$ and then proceeds along one of two possible paths: The first path involves assuming a functional form for~$f_X(x)$ and then using the data~$x_i$ to estimate the parameters of this distribution and the associated estimate~$\hat{f}_X(x)$. For example, if one assumes that the samples~$x_i$ come from a normal distribution, the mean~$\hat{\mu}$ and standard deviation~$\hat{\sigma}$ are estimated from the data and one uses the normal distribution~${\cal N}(\hat{\mu},\hat{\sigma})$ in the betting analysis to follow. The second possibility is that no constraints are imposed upon the form of~$f_X$ and one simply works with an empirical approximation~$\hat{f}_X(x)$ for the true PDF~$f_X(x)$. This empirical Probability Mass Function (PMF) is given by the sum of impulses
$$
\hat{f}_X(x) = \frac{1}{m}\sum_{i = 1}^{m} \delta(x -x_i).
$$
In this case, when Kelly betting is considered, $\hat{f}_X(x)$ is used as input to the optimization of~$g(K)$ and a maximizer, call it~$\hat{K}^*$, is used as the betting fraction. For the background probability theory underlying the analysis to follow, the reader is referred to~\cite{Gubner_2006}.  

\vskip 2mm

Given the scenario above, the following questions present themselves: If we base our betting fraction~$\hat{K}$ on~$\hat{f}_X$ rather than~$f_X$, how will the optimum~$\hat{K}^*$ compare with the ``true'' optimum~$K^*$? What sample size~$m$ is needed so that the empirically-based optimal performance is acceptably close to the true optimum? Perhaps the simplest possible illustration of these ideas is obtained by considering $X$ to be a scalar corresponding to the outcome of repeated flipping of a biased coin with probability of heads being~$p > 1/2$. Assuming an even-money bet, we take $X=1$ for heads and $X=-1$ for tails. Then, if one has a perfect knowledge of~$p$, it is readily verified that~$g(K)$ is maximized via
$
K^* = 2p-1.
$
On the other hand, if the Kelly bets are being derived from empirical samples~$x_1,x_2,\dots,x_m$ in~~$\{-1,1\}$ with~$x_i = 1$ being the return for ``heads,"  then the sample mean
$$
\hat{p} = \frac{1}{m} \sum_{i = 1}^m \max \{ x_i, 0\}
$$ 
is used as input to the analysis and one obtains 
$$
\hat{K}^* = \max\{2\hat{p}-1,0\}
$$ 
as the optimal betting fraction.

\vskip 3.5mm

\section{HOW OVERLY CONSERVATIVE BETS ARISE}
Beginning with an empirically derived PMF as described above, our first objective in this section is as follows: We describe the key ideas driving many scenarios where the Kelly bettor who uses ``pure theory'' in lieu of empirical data may reach a conclusion about the optimal bet size which entirely contradicts common sense real-world considerations. That is, we describe a scenario which demonstrates how formal application of the Kelly theory can lead to a bet size which is far smaller than merited by analysis of risk versus return. Our second objective is to provide a realistic numerical example showing that this pathology which we describe is realizable using real data. To this end, we consider a scenario involving samples drawn from a normal distribution.

\vskip 2mm

{\bf Pathology Explained for a Toy Example}: We consider one of the simplest possible Kelly betting problems. It is described by a Bernoulli random variable~$X$ whose PMF is given as follows:~$P(X = 1) = 1-\varepsilon$ and $P(X = -x_0) = \varepsilon$ where~$x_0 \gg 1$, and
$$
0<\varepsilon < \frac{1}{1+x_0}.
$$
For this simple scenario, the Kelly betting problem is easily solved via existing literature. For the sake of completeness, we describe the solution. Indeed, we initially hold~$\varepsilon$ and~$x_0$ fixed and later consider the consequence of varying these parameters. We first compute
\begin{align*}
	g(K) 
	&=  \mathbb{E}「[\log(1 + KX)]\\
	 &=  (1-\varepsilon)\log(1+K) + \varepsilon\log(1-Kx_0)
\end{align*}
and note that this function is readily maximized with respect to~$K$ using ordinary calculus.  Via a lengthy but straightforward calculation, we obtain the optimal Kelly fraction~\mbox{$K = K^*$} with
$$
K^* = \frac{1 - \varepsilon(1+x_0)}{x_0}
$$
which is readily verified to satisfy
$$
0 < K^* < \frac{1}{x_0}. 
$$
This is consistent with the observation that~$K \geq 1/x_0$ leads to~$\log(1 -Kx_0) = -\infty$ irrespective of the size of~$\varepsilon$. Now, the key point to note is the following: {\it No matter how small~$\varepsilon$ is, the size of~$K^*$ is limited by~$1/x_0$.} In other words, even when the risk~$\varepsilon$ of losing becomes negligible, for the Kelly bettor using this theoretical model, the size of the bet will be inappropriately small. For example, with~$x_0 = 100$, no matter how small~$\varepsilon$ is, the betting fraction~$K$ can never be more than~$1$\% of the account value. 
In summary, when situations arise with common sense dictating that one should wager almost all of one's account, the formal Kelly theory forces the betting fraction to be far too small; i.e., an overly conservative bet results.

\vskip 2mm

To complete the arguments related to this toy example, we now imagine a ``practitioner'' who is enamored with Kelly theory but distrusting of a theoretical model. Suppose further that empirical data for the random variable~$X$ above is available, perhaps in limited supply. In this case, per the discussion in Section 2, this bettor collects~$m$ data points, generates an empirical PMF, and then, based on this estimated distribution, determines the optimal bet. What will happen when~$\varepsilon$ is extremely small? Clearly, without~$m$ being unacceptably large, it is virtually certain that the bettor will see~$x_i = 1$ for~$i = 1,2,\ldots,m$. Hence, the empirically derived PMF for the estimated random variable~$\hat{X}$ is trivially described. Namely,~$\hat{X} = 1$ with probability one and the resulting expected log-growth maximizer, namely~$\hat{K}^* = 1$ is more consistent with the common sense maxim: ``When conditions are right, bet the farm.''

\vskip 2mm

The arguments above are not intended to be entirely rigorous because the role of the sample size~$m$ has not really been considered. To tighten up the arguments above, we note the following: In practice, there is a limitation on~$m$, say~\mbox{$m \leq M$}, which can arise for various reasons. For example, if~$X(k)$ represents daily returns on a stock, then it would typically be the case that~$m$ is strongly limited because the underlying assumption of independent and identically distributed returns becomes questionable when~$m$ is too large. For example, many traders do not use large~$M$ in the belief that larger~$M$-values require processing of ``old data'' which may not reflect current market conditions. For the case of the random variable~$X$ in the toy example above, we can ask: What is the probability, call it~$p_{bad}$, that the practitioner will see a ``bad'' sample; i.e.,~$x_i = -x_0$ for some~$i \leq M$. For this simple problem, we obtain
\[
p_{bad} = 1- (1-\varepsilon)^M.
\] 
Thus, if~$\varepsilon = 0.001$ and~$M=50$, then we obtain~$p_{bad} \approx 0.05$ and if~$\varepsilon = 0.0001$,~$p_{bad} \approx 0.005$. Note that if such a bad sample is ``seen,'' the behavior of the practitioner becomes similar to that of the theoretical Kelly bettor.

\vskip 2mm

\begin{figure}[htbp]
	\centering
	\graphicspath{{Figs/}}
	\setlength{\abovecaptionskip}{0.1 pt plus 0pt minus 0pt}
	\includegraphics[width=0.47\textwidth]{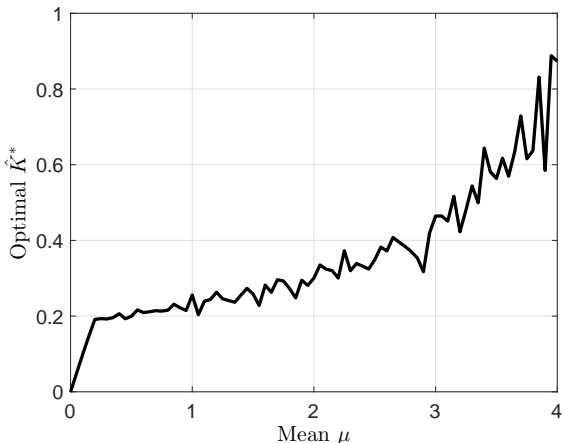}
	\figcaption{Optimal Kelly Fraction $\hat{K}^*$ Versus $\mu$ }
	\label{fig.K_vs_mu}
\end{figure}

{\bf A More Realistic Example}: To study the issue of conservatism using realistic data, we consider a family of random variables each of which is governed by the normal distribution. Each of these random variables has fixed standard deviation~$\sigma =1$. However, the members of this family are differentiated by their means. We consider means~\mbox{$0 \leq \mu \leq 4$}. For each value of~$\mu$ in this range, we let~$X_{\mu}$ denote the random variable of interest and construct an empirical probability mass function drawing~$m=1,000,000$ samples. Next, for each~$\mu$, we find the optimal Kelly fraction, call it~$\hat{K}^* = \hat{K}^*(\mu)$; see Figure~\ref{fig.K_vs_mu} where this function is plotted. Looking at the plot, we now argue that this result is consistent with common sense considerations.  Indeed, when~$\mu$ is at the low end of the range, it is no surprise to see that~$\hat{K}^*(\mu)$ is small because the probability of~$X_\mu < 0$ is significant. For example, when~$\mu = 1$, the optimum is to wager about~$20\%$ of one's wealth on each bet. Similarly, when~$\mu$ is at the high end of the range, we see that $\hat{K}^*(\mu)$ is large because the probability of~$X_\mu < 0$ becomes small. For example, when~$\mu = 4$, the optimum is to wager about~$90\%$ of one's wealth on each bet because the chance of losing is vanishingly small.

\vskip 2mm

In the next section, we see that this analysis using real data is entirely at odds with a purely theoretical analysis. In this regard, when the analysis in the section to follow is used to analyze the random variables~$X_\mu$, one ends up with optimal betting fraction~$K^* = 0$; i.e., no betting at all is dictated. 
To conclude this section, we note the following: The fact that our data-based analysis above was carried out with fixed~$\sigma$ is not critical to the conclusions we reached. More generally, when $X$ is governed by normal distribution~$\mathcal{N}(\mu, \sigma)$, the Kelly theory suggests no betting regardless of the relative sizes of the mean~$\mu$ and standard deviation~$\sigma$. 

\section{RESTRICTED-BETTING: THE SCALAR CASE}
In this section, we present an analysis regarding the motivating examples in the preceding section. In rough terms, for a scalar random variable~$X$, we see that the minimum and maximum values of points~$x$ in the support lead to fundamental restrictions on the size of the bet allowed by the Kelly theory --- the larger these values, the smaller the Kelly fraction is forced to be. Moreover, this restriction holds true whether the probability of these maximal deviations is significant or not.

\vskip 2mm

Since the key ideas driving the analysis to follow are most simply understood when $X$ is a scalar random variable, we first consider this case. To begin, suppose~$x_0 < 0$ is a point in the support set of~$X$. Then, to avoid~$g(K) = -\infty$, Kelly theory forces the betting fraction to satisfy~$K \le -1/x_0$. This holds true even when the probability that~$X$ gets close to~$x_0$ is vanishingly small. Similarly, for a point~$x_0 > 0$ in the support, similar reasoning forces~$K \ge -1/x_0$. As a result of this aspect of the theory, many bets which are ``excellent'' from a common sense point of view lead to unduly small bets. We note that this is consistent with the examples in Section~3. To summarize, in the Kelly theory, large values of~$X$, whether rare or not, lead to dramatic restrictions in the bet size.

\vskip 2mm

In the lemma below, we formalize the ideas above. An extreme case of the result occurs when the support of~$X$ is the entire real line; e.g., suppose~$X$ is normally distributed. For such cases, as seen below,~$K = 0$ is forced. That is, no betting is allowed. This result holds true regardless of the relative sizes of the mean~$\mu$ and standard deviation~$\sigma$. We note that this outcome of Kelly theory is clearly at odds with practical considerations. Even when the ratio~$\mu/\sigma$ is very large, synonymous with an excellent bet, the theory nevertheless forces~$K = 0$. The lemma below is a special case of the Restricted Betting Theorem given in the next section. Accordingly, its proof is deferred until then.

\vskip 2mm

{\bf Scalar Betting Lemma}: {\it Let $X$ be a random variable with~$\mathbb{E}[|X|]<\infty$,  probability density function~$f_X(x)$ and support~${\cal X}$ with extremes 
\[
X_{\min} \doteq \inf \{x: x \in \mathcal{X} \} 
\text{ \; and \; } 
X_{\max} \doteq \sup \{x: x \in \mathcal{X} \}
\]
satisfying $X_{\min}<0 $ and $ X_{\max} >0$. Then any optimizing Kelly fraction~$K$ maximizing~$g(K)$ satisfies the interval confinement condition
$$
K \in [-1/X_{\max},-1/X_{\min}].
$$
}{\bf Remarks}: $(i)$ Consistent with the remarks prior to the  statement of the lemma,~$K^* \leq 0$ when~$X_{\min} = -\infty$ and~$K^* \geq 0$ when~$X_{\max} = +\infty$.  It follows that~$K = 0$ is forced.  In other words, the best bet is no bet at all.

\vskip 2mm

$(ii)$ The lemma says that an optimal $K$ must lie in the confinement interval, but we do not expect {\it every} $K$ in the interval to be optimal. Surprisingly, there may exist some~$K$ in the confinement interval that are ``infinitely bad,'' i.e.,~$g(K)=-\infty$, as shown in the following example.

\vskip 2mm

{\bf Example}: We provide an example of a random variable~$X$ and a constant~$K > 0$ satisfying the confinement condition above
but having the property that~$g(K) = -\infty.$ Indeed, let~$0 < K < 1$ be arbitrary and held fixed in the calculations to follow. We now consider a random variable~$X$ which  is constructed as follows. Let
\begin{eqnarray*}
	\theta  \doteq  \frac{1}{2} + \sum_{k=1}^\infty \frac{1}{k^2}
	  =  \frac{1}{2} + \frac{\pi^2}{6},
\end{eqnarray*}
take $X =x_0 =1$ with probability~$p_0=1/(2 \theta)$, and for~\mbox{$k\ge 1$}, take
$
X = x_k \doteq (e^{-k}-1)/K
$
with probability~$p_k \doteq {1}/({k^2 \theta}).$ Note that the definition of $\theta$ above assures that the $p_k$ define a probability mass function;~i.e.,~$p_k \ge 0$ and~$\sum_{k=0}^\infty p_k =1.$
Now, for this random variable, we have~$X_{\min} = -{1}/{K},$
and~\mbox{$X_{\max} = 1.$}
Furthermore, since~\mbox{$0 < K < 1$}, the interval confinement condition is satisfied. To complete the analysis, it remains to show that~$g(K) = -\infty$. Indeed, we calculate
\begin{align*}
g\left( K \right)
&= \mathbb{E}[\log \left( {1 + KX} \right)]\\
& = \sum\limits_{k = 0}^\infty  {\log \left( {1 + K{x_k}} \right){p_k}} \\
& = \log \left( {1 + K{x_0}} \right){p_0} + \sum\limits_{k = 1}^\infty  {\log \left( {1 + K{x_k}} \right){p_k}} \\
& = \frac{1}{{2\theta }}\log \left( {1 + K} \right) + \frac{1}{\theta }\sum\limits_{k = 1}^\infty  {\frac{1}{{{k^2}}}\log \left( {1 + K{x_k}} \right)} \\
& = \frac{1}{{2\theta }}\log \left( {1 + K} \right) - \frac{1}{\theta }\sum\limits_{k = 1}^\infty  {\frac{1}{k}}  =  - \infty.
\end{align*}

\vskip 1mm

\section{THE RESTRICTED BETTING THEOREM}
Recalling the interval confinement condition introduced for a scalar random variable,  this section provides a generalization of this result which holds for an~\mbox{$n$-dimensional} random vector~$X$ whose support set~${\cal X}$ can be rather arbitrary. This support set is allowed to be unbounded so that we capture the no-betting result given for~$X$ being a scalar.  To obtain the theorem below, we make use of the classical support function which is heavily used in convex analysis;~e.g., see~\cite{Rockafellar_1996}.

\vskip 2mm

Indeed, given a set~${\cal X} \subseteq {\bf R}^n$, the {\it support function} on~${\cal X}$ is the mapping~$h: {\bf R}^n \rightarrow {\bf R} \cup \{+\infty \}$ defined as follows: For~$y \in {\bf R}^n$,
$$
h_{\cal X}(y) \doteq \sup_{x \in \cal X} y^Tx.
$$
After establishing the theorem below, we consider a number of special cases to show that there are large classes of Kelly betting problems for which checking for satisfaction of the conditions is highly tractable.

\vskip 2mm

{\bf The Restricted Betting Theorem}: {\it Given  an~\mbox{$n$-dimensional} random vector $X$ with PDF~$f_X$, support~${\cal X}$, and~\mbox{$\mathbb{E}[\| X \|] < \infty$}, any optimizing Kelly fraction vector~$K$ satisfies the condition
	$$
	h_{\cal X}(-K) \leq 1.
	$$
	Furthermore, whether~${\cal X}$ is convex or not, the set
	$$
	{\cal K} \doteq \{K: h_{\cal X}(-K) \leq 1\}
	$$
	is convex and closed.
}

\vskip 2mm

{\bf Proof}: In the arguments to follow, we work with the extended logarithmic function which takes value~$\log(x) = -\infty$ for~$x \le 0$. Proceeding by contradiction, suppose~$K$ is optimal but fails to satisfy the support function condition above. Then
$$
\sup_{x \in \cal X} [-K]^T x > 1.
$$
Equivalently, there exists some~$x^K \in {\cal X}$ such that~$-K^Tx^K > 1.$
Hence $1 + K^Tx^K < 0.$
Now noting that~$1 + K^Tx$ is continuous in~$x$ and that~$x^K$ is in the support, there exists a suitably small neighborhood of~$x^K$, call it~${\cal N}(x^K)$, such that $1 + K^Tx < 0$
for~$x \in  {\cal N}(x^K)$ and
$$
P(X \in {\cal N}(x^K)) > 0.
$$
We now claim that the existence of such a neighborhood implies that $g(K) = -\infty$. Indeed, to prove this, we first observe that
\begin{align*}
g(K) 
&=\mathbb{E}[ \log(1+K^TX)]\\
&= \int \log(1+K^T x) f_X(x)dx \\
&=  \int \limits_{  1 + K^T x \le 0  }^{} \hskip -4mm {\log (1 + {K^T}x){f_X}\left( x \right)dx}  \\
& \;\;\;\;\;\;\;\;\; + \int \limits_{ 1 + {K^T}x > 0  }^{} \hskip -4mm {\log (1 + {K^T}x){f_X}\left( x \right)dx}. 
\end{align*}
Using the property of logarithmic function that
$$
\log(1+K^T x) \le |K^T x|
$$
for all $x$ satisfying $1+K^T x >0$, we obtain an upper bound for $g(K)$. That is,
{\small \begin{align*}
g(K) 
&\le \int \limits_{1 + {K^T}x \le 0}^{} \hskip -4mm  {\log (1 + {K^T}x){f_X}\left( x \right)dx}   + \int \limits_{ 1 + {K^T}x > 0 }^{} \hskip -4mm {\left| {{K^T}x} \right|{f_X}\left( x \right)dx}\\
&\le \int\limits_{1 + {K^T}x \le 0}^{} \hskip -4mm{\log (1 + {K^T}x){f_X}\left( x \right)dx}   + \int_{}^{} {\left\| K \right\|\left\| x \right\|} {f_X}\left( x \right)dx\\
 &\le \int\limits_{1 + {K^T}x \leq 0}^{} \hskip -4mm{\log (1 + {K^T}x){f_X}\left( x \right) dx}   + \| K \| \; \mathbb{E}[\| X \|].
\end{align*}
}Since $ \mathbb{E}[\| X \|] < \infty$, it suffices to show that the integral above has value $-\infty$. Indeed, beginning with the fact that
$$
P(X \in {\cal N}(x^K)) > 0
$$
and noting that~${\cal N}(x^K) \subseteq \{x: 1+K^T x\le 0\}$, the density function~$f_X$ must assign positive probability  to the set~\mbox{$\{x: 1+K^T x \leq 0 \}$}. Furthermore, since~\mbox{$\log(1+K^Tx) = -\infty$} for~$x$ satisfying~\mbox{$1+K^Tx \le 0$}, it follows that
\[
\int \limits_{1 + {K^T}x \le 0}^{} \hskip -4mm{\log (1 + {K^T}x){f_X}\left( x \right)dx}  = -\infty
\]and we conclude that
$
g(K) = -\infty
$ as required. 

\vskip 2mm

To complete the proof, we establish closedness and convexity of~${\cal K}$ using a rather standard convex analysis argument: Indeed, for each fixed~$x \in {\cal X}$, we define the linear function~$L_x(K) \doteq -K^Tx$ and associated set
$$
{\cal K}_x \doteq \{K: L_x(K) \leq 1\}.
$$
Note, that~${\cal K}_x$, being a halfspace, is a closed convex set. Now, using the definition of the support function, it follows that
$$
{\cal K} = \bigcap_{x \in \mathcal{X}}^{} {\cal K}_x.
$$
Hence, since~${\cal K}$ is the intersection of an indexed collection of closed convex sets, it is also closed and convex. $\square$

\vskip 2mm

{\bf Scalar Result as a Special Case}: To see that the Scalar Betting Lemma in Section 5 is a special case of the above, we recall notation $X_{\min}$ and~$X_{\max}$ and assume~\mbox{$X_{\min} < 0$} and~$X_{\max} > 0$ as in the earlier sections. Now, for~$K > 0$, the support function in the theorem above becomes~\mbox{$h_{\cal X}(-K) = -KX_{\min}$}
and for~$K < 0$, it becomes~$h_{\cal X}(-K) = -KX_{\max}.$
Hence the requirement of the theorem~$h_{\cal X}(-K) \leq 1$ leads to the interval confinement condition of the lemma.

\vskip 2mm

{\bf Hypercube Support Set}: One $n$-dimensional generalization of the scalar situation above is obtained when the convex hull of the support of~$X$, $\mbox{conv}{\cal X}$, is a hypercube. Suppose this hypercube has center~$x^0$ and components~$x_i$ satisfying~\mbox{$|x_i - x^0_i| \leq \delta_i$}
where~$\delta_i > 0$ for~$i = 1,2,\ldots,n$. Then using a basic fact about support functions, see~\cite[p.~269]{Witsenhausen_1980}, that
$
h_{\cal X}(y) = h_{\mbox{conv}{\cal X}}(y)
$
for all~$y \in {\bf R}^n$, a straightforward calculation leads to
$$
h_{\cal X}(-K) =  \sum_{i=1}^{n}|K_i|\delta_i -\sum_{i=1}^{n}K_ix_i^0 .
$$
Hence, application of the theorem leads to the requirement that 
any optimizing Kelly fraction vector~$K$ satisfies the condition
$$
\sum\limits_{i = 1}^n | {K_i}|{\delta _i} - \sum\limits_{i = 1}^n {{K_i}} x_i^0 \le 1.
$$

\vskip 2mm

\begin{figure}[htbp]
	\centering
	\graphicspath{{figs/}}
	\setlength{\abovecaptionskip}{0.1 pt plus 0pt minus 0pt}
	\includegraphics[width=0.40\textwidth]{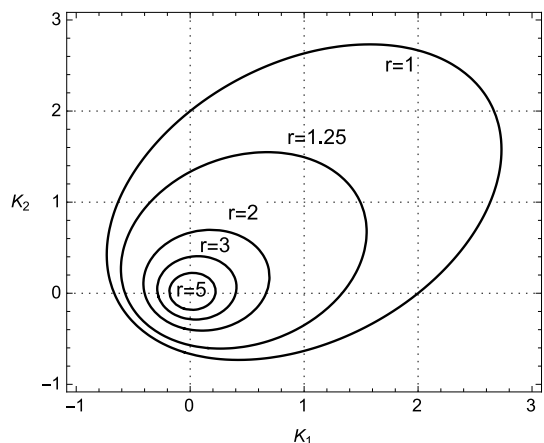}
	\figcaption{Constraint Sets $\mathcal{K}_r$ for Optimal Fraction $K$}
	\label{fig:HyperSphere}
\end{figure}

{\bf Hypersphere Support  Set}: As a final example, suppose the convex hull of the support set~${\cal X}$ is a hypersphere in~${\bf R}^n$ with description
$
\| x - x^0 \| \leq r
$
with euclidean norm used above, center~$x^0$,  and radius~$r > 0$. Then using an argument which is similar to that used for the hypercube example above, we can easily show that any optimizer~$K$ must satisfy
$$
 r \| K \| - K^Tx^0  \le  1.
$$
We note that the constraint sets
\[
\mathcal{K}_r \doteq \{K: r \|K \| - K^Tx^0  \le  1 \}
\]
are nested. That is,  if radii $ r_1 \le r_2$, then the set $\mathcal{K}_{r_2} \subseteq \mathcal{K}_{r_1}$. In Figure~\ref{fig:HyperSphere}, these sets are depicted for $x^0 = (1/2,1/2)$ and various radii $r_1=1$, $r_2=1.25$, $r_3=2$, $r_4=3$ and~$r_5=5$. 

\vskip 3.5mm
\section{ EXAMPLE INVOLVING HIGH-FREQUENCY} \label{EX:High-freq}
Thus far, our analysis of the restricted betting phenomenon has included no consideration of the frequency with which wagers are being made. In this regard, we imagine the frequency of betting to be so high as to make it seem ``reasonable'' for the theoretician to use a continuous-time stochastic model to determine the optimal betting fraction. The question we consider is as follows: For the high-frequency case with sufficiently many samples being used to construct the empirical distribution, is there still a disparity between theory and practice? That is, is it still the case that the theoretical solution can end up being far too conservative? In~\cite{kuhn_and_luenberger}, an issue with rather similar flavor is considered in the context of portfolio optimization and the analysis given is much more abstract than that given below. Here we consider a concrete example and provide no significant result of general import. Our main objective is to raise issues for future research.

\vskip 2mm

\begin{figure}[htbp]
	\centering
	\graphicspath{{figs/}}
	\setlength{\abovecaptionskip}{0.1 pt plus 0pt minus 0pt}
	\includegraphics[width=0.475 \textwidth]{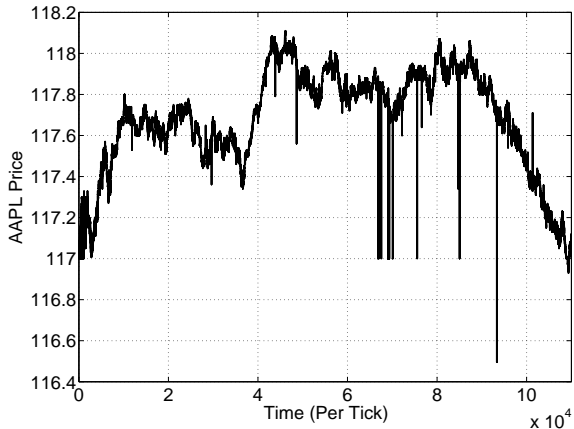}
	\figcaption{AAPL Tick-by-Tick Price of Trade}
	\label{fig:AAPL_Prices}
\end{figure}

Indeed, we begin with high-frequency historical intra-day tick data for APPLE (ticker AAPL). Each ``tick'' corresponds to a new stock price~$S(k)$ and the time between arrivals of ticks is estimated on average to be about one tenth of a second. This stock-price data is plotted in Figure~\ref{fig:AAPL_Prices} for the period~9:30:00~am to~2:13:47~pm on December~2,~2015. During this period, we have $m = 110,000$ ticks. The first step in our analysis was to  
use the time series prices~$S(k)$ to calculate the corresponding returns
$$
X(k) = \frac{S(k+1) - S(k)}{S(k)}.
$$
Given the small time between consecutive ticks, a large percentage of the~$X(k)$ turn out to be zero;~i.e., the price did not change from~$k$ to~$k+1$.
 In addition, the smallness of the inter-tick times leads to the remaining probability masses largely concentrated between~$x = -0.0002$ and~\mbox{$x = 0.0002$}
and the data leads to~\mbox{$X_{\min} \approx -0.01 \approx -X_{\max}$}. Thus, the Restricted Betting Theorem forces the approximate bound~\mbox{$-100 \leq  K \leq 100$} which is not really meaningful since brokerage requirements typically limit~$ |K| \leq 2$. Based on the empirical data,  we plotted~$g(K)$ and obtained the optimal betting fraction~\mbox{$\hat{K}^* \approx 0.824$}. Interestingly, although the price has no obvious ``bullish'' pattern, we see that the theory leads to a rather aggressive bet size which is more than~$80 \%$ of one's wealth. 
In contrast, if we assume that the data for this example, comes from a discrete-time Geometric Brownian Motion  with this same mean and variance, we obtain~$K^* = 0$ by the Restricted Betting Theorem. 

\vskip 2mm

It is interesting to note that other methods in the literature which might be used for the same problem lead to optimal~$K$-values which are remarkably close to~$\hat{K}^* \approx 0.824$ obtained above. For example using estimated mean~\mbox{$\hat{\mu} \approx 1.628 \times 10^{-8}$} and standard deviation~\mbox{$\hat{\sigma} \approx 1.405 \times 10^{-4}$} as the basis for a continuous time Geometric Brownian Motion model, the analysis in~\cite{merton} involves optimizing the expected value having combination of consumption utility and logarithm of terminal wealth. As the consumption weighting tends to zero, the optimal fraction tends to~\mbox{$ K^* = {\hat{\mu}} /{\hat {\sigma}^2} \approx 0.825.$} The same result is obtained in~\cite{Thorp_2006} using the same expected logarithmic growth criterion and assuming a stochastic process model with bounded returns~$X(k)$ with mean $\hat{\mu}$ and standard deviation~$\hat{\sigma}$.

%

\vskip 2mm

\section{CONCLUSION AND FUTURE WORK}
In this paper, we considered a random vector~$X$ and compared the size of Kelly bets which are derived using a purely theoretical probability distribution versus those which are obtained from its empirically-obtained counterpart. In making this comparison, the support set~${\cal X}$ for~$X$ was seen to play a crucial role. As seen in the Restricted Betting Theorem in Section~6, when the logarithmic growth function~$g(K)$ is maximized, this set~${\cal X}$ can lead to ``unreasonable'' restrictions on the optimal betting fraction~$K^*$. By this we mean roughly the following: Possible outcomes for~$X$ which are ``large'' can lead to the possibility that extremely attractive betting opportunities are rejected. On the other hand, when betting is based on an empirically derived distribution for~$X$, it is likely that such rare events will not be reflected in the resulting probability mass function. The bet size which results will be more in line with common sense.

\vskip 2mm

These results open the door to a new line of research which might be appropriately called ``data-driven Kelly betting.'' In such an empirical framework, new problems involving the sample size~$m$ will be of fundamental importance. Given that many betting processes involve non-stationary stochastic processes, there is typically a bound~$m \leq M$ which must be respected when deriving the empirical distribution. That is, when the analysis involves sequential betting based on i.i.d. random variables, the use of ``untrustworthy old data'' from far in the past may be inappropriate to use.

\vskip 2mm

A second important future research direction involves extension of Kelly-based analysis to problems involving the betting frequency. This topic, touched upon in Section~7, does not appear to have been heavily considered in the literature; e.g., see~\cite{kuhn_and_luenberger} for results available to date. In this setting, many new modeling and analysis questions arise involving what betting frequencies are available and the model of the random variable~$X$ changes as a function of frequency. For example, if even-money coin flips are carried out at some frequency~$f$, the model for~$X$ does not change from bet to bet; i.e., the bet is independent of frequency. On the other hand, if~$X$ corresponds to the return on a stock based on sampling of a continuous-time Brownian motion, appropriate scaling of the mean and variance become important issues as the frequency increases. 


\addtolength{\textheight}{-3cm}   



\begin{thebibliography}{99}
	
	
	\bibitem{Kelly_1956}
	J. L. Kelly, ``A New Interpretation of Information Rate," {\it Bell System Technical Journal,}  pp. 917-926,~1956.
	
	\bibitem{merton}	
	R. C. Merton, ``Lifetime Portfolio Selection Under Uncertainty: The Continuous-Time Case," {\it Review of Economics and Statistics,}~vol.~51,~pp.~247-257,~1969.
		
	\bibitem{Hakansson_1971}
	N. H. Hakansson, ``On Optimal Myopic Portfolio Policies With and Without Serial Correlation of Yields," {\it Journal of Business,}~vol.~44, \mbox{pp.~324-334},~1971.
		
	\bibitem{Finkelstein_Whiteley_1981}
	M. Finkelstein and R. Whitley, ``Optimal Strategies for Repeated Games," {\it Advanced Applied Probability},~vol. 13,~pp. 415-428,~1981.
				
	\bibitem{Algoet_Cover_1988}
	P. H. Algoet and T. M. Cover, ``Asymptotic Optimality and Asymptotic Equipartition Properties of Log-Optimum Investment," {\it The Annals of Probability,}~vol.~16,~\mbox{pp.~876-898},~1988.
	
	\bibitem{Maclean_Ziemba_Blazenko_1992}
	L. C. Maclean, W. T. Ziemba and G. Blazenko ``Growth Versus Security in Dynamic Investment Analysis," {\it Management Science,} vol.~38, pp. 1562-1585,~1992.
		
	\bibitem{Maclean_Ziemba_1999}
	L. C. Maclean and W. T. Ziemba ``Growth Versus Security Tradeoffs in Dynamic Investment Analysis," {\it Annals of Operations Research,} vol.~85, pp. 193-227,~1999.
			
	\bibitem{Thorp_2006}
	E. O. Thorp, ``The Kelly Criterion in Blackjack Sports Betting and The Stock Market," {\it Handbook of Asset and Liability Management: Theory and Methodology,}~vol. 1,~pp. 385-428, Elsevier Science,~2006.
		
	\bibitem{Maclean_Thorp_Ziemba_2010}
	L. C. Maclean, E. O. Thorp, and W. T. Ziemba ``Long-term Capital Growth: The Good and Bad Properties of The Kelly and Fractional Kelly Capital Growth Criteria," {\it Quantitative Finance,}~vol.~10,~\mbox{pp. 681-687},~2010.

	
	\bibitem{Davis_Lleo_2010}
	M. Davis and S. Lleo, ``Fractional Kelly Strategies for Benchmarked Asset Management," in L. C. MacLean, E. O. Thorp, and W. T. Ziemba, {\it The Kelly Capital Growth Investment Criterion: Theory and Practice,} World Scientific,~\mbox{pp. 385-407},~2010.
	
				
	\bibitem{MacLean_Thorp_Ziemba_2011}
	L. C. MacLean, E. O. Thorp, and W. T. Ziemba, {\it The Kelly Capital Growth Investment Criterion: Theory and Practice,} World Scientific Publishing Company,~2011.
	

	\bibitem{Rising_Wyner_2012}
	J. K. Rising and A. J. Wyner, ``Partial Kelly Portfolios and Shrinkage Estimators," {\it Proceedings of IEEE International Symposium on Information Theory}, pp. 1618-1622,~2012.
		
	\bibitem{Nekrasov_2014}
	V. Nekrasov, ``Kelly Criterion for Multivariate Portfolios: A Model-Free Approach," {\it Social Science Research Network Electronic Journal},~2014.

	\bibitem{Ziemba_2015}
	W. T. Ziemba, ``Response to Paul A Samuelson Letters and Papers on the Kelly Capital
	Growth Investment Strategy,'' {\it Journal of Portfolio Management}, vol.~42,~pp.~153-167,~2015	
	
	\bibitem{Barmish_Hsieh_2015}
	C. H. Hsieh and B. R. Barmish, ``On Kelly Betting: Some Limitations," {\it Proceedings of the Annual Allerton Conference on Communication, Control, and Computing,} pp. 165-172,~2015.


	
	\bibitem{Gubner_2006}
	J. A. Gubner, {\it Probability and Random Processes for Electrical and Computer Engineers,} Cambridge University Press,~2006.

	\bibitem{Rockafellar_1996}
	R. T. Rockafellar, {\it Convex Analysis,} Princeton University Press,~1996.

	\bibitem{Witsenhausen_1980}
	H. S. Witsenhausen, ``Some Aspects of Convexity Useful in Information Theory," {\it IEEE Transactions of Information Theory,}~vol.~26,~pp.~265-271,~1980.

	\bibitem{kuhn_and_luenberger}	
	D. Kuhn and D. G. Luenberger, ``Analysis of the Rebalancing Frequency in Log-optimal Portfolio Selection," {\it Quantitative Finance,}~vol.~10,~pp.~221-224,~2010.	
	
	


\end{thebibliography}
\end{document}